# Angular momentum of circularly polarized light in dielectric media

**Masud Mansuripur**

*College of Optical Sciences, The University of Arizona, Tucson, Arizona 85721*
*masud@optics.arizona.edu*

**Abstract**: A circularly polarized plane-wave is known to have no angular momentum when examined through Maxwell's equations. This, however, contradicts the experimentally observed facts, where finite segments of plane waves are known to be capable of imparting angular momentum to birefringent platelets. Using a superposition of four plane-waves propagating at slightly different angles to a common direction, we derive an expression for the angular momentum density of a single plane-wave in the limit when the propagation directions of the four beams come into alignment. We proceed to use this four-beam technique to analyze the conservation of angular momentum when a plane-wave enters a dielectric slab from the free space. The angular momentum of the beam is shown to decrease upon entering the dielectric medium, by virtue of the fact that the incident beam exerts a torque on the slab surface at the point of entry. When the beam leaves the slab, it imparts an equal but opposite torque to the exit facet, thus recovering its initial angular momentum upon re-emerging into the free-space. Along the way, we derive an expression for the outward-directed force of a normally incident, finite-diameter beam on a dielectric surface; the possible relationship between this force and the experimentally observed bulging of a liquid surface under intense illumination is explored.

**OCIS codes**: (260.2110) Electromagnetic theory; (140.7010) Trapping.

## 1. Introduction

The Poynting vector $\mathbf{S}$ of a circularly polarized plane-wave propagating in free-space along the $z$-axis is a constant vector parallel to $z$. The linear momentum density of the field, $\mathbf{p} = \mathbf{S}/c^2$, is thus aligned with the $z$-axis and, as such, cannot give rise to any $z$-oriented angular momentum $\mathbf{L} = \mathbf{r} \times \mathbf{p}$. Experimentally, however, it is well known that a circularly polarized plane-wave, intercepted on a finite area, contains an angular momentum of $\hbar$ per photon [1-3]. To resolve this discrepancy, we note that while $p_x$ and $p_y$ are zero for a plane wave, $\mathbf{r}$ can extend all the way to infinity, yielding the product of 0 and $\infty$ indeterminate. We thus resort to a limiting procedure whereby a number of plane-waves are made to propagate at a slight angle $\theta$ to the $z$-axis. The superposition of these plane-waves creates well-defined interference fringes, each of which occupies a finite area in any cross-sectional plane of the beam. We confine our attention to one such fringe, compute the angular momentum density $\mathbf{L}$ averaged over the area of the fringe, then bring the constituent plane-waves into alignment by allowing their deviation angle $\theta$ to approach zero. In this way the fringe area extends to infinity, the superposed plane-waves begin to resemble a single plane-wave propagating along $z$, and the $0 \times \infty$ ambiguity is resolved in a limiting procedure that yields a definite value for $\mathbf{L}$, consistent with experimental observations.

Although alternative methods exist that address the aforementioned problem of circularly polarized plane-waves [4], our four-wave technique has the advantage of simplicity, enabling one to explore more complex issues, such as the controversial problem of the angular momentum of light in dielectric media. Recent years have witnessed a growing interest in both experimental and theoretical aspects of the optical angular momentum, with applications emerging in areas such as optical spanners and micro-manipulators [5,6]. Loudon [7] has given a comprehensive quantum analysis of the angular momentum of light in dielectric media, and has rightly emphasized the importance of direct calculations of force and torque by starting from the Lorentz law [8]. While we agree with many aspects of Loudon's theoretical treatment, we differ on the correct form of the Lorentz law as applied to dense dielectric materials. Our recent publications [9-12] give extensive accounts of the classical electromagnetic theory applied to the problem of the linear momentum of light in dielectrics. The present paper extends these results to the case of optical angular momentum.

In Section 2 we introduce the concept of four-wave superposition, and explore the properties of such beams in free-space. Section 3 examines the properties of four-wave beams inside dielectric media. The force and torque of a four-wave beam at the interface between the free-space and a semi-infinite dielectric are calculated in Section 4, followed by a derivation in Section 5 of the beam's force and torque exerted on the bulk of the semi-infinite medium. A summary of the results and a discussion that highlights our differences with the results of Loudon [7] are given in Section 6.

## 2. Superposition of four plane-waves

The electric field $\mathbf{E}$ and the magnetic field $\mathbf{H}$ of a circularly-polarized, monochromatic plane-wave propagating in free space at angle $\theta$ to the $z$-axis may be expressed as follows:

$$\mathbf{E}_1(x, y, z) = E_o (\cos\theta\, \hat{\mathbf{x}} + i\hat{\mathbf{y}} - \sin\theta\, \hat{\mathbf{z}})\, \exp[ik_o(x\sin\theta + z\cos\theta)], \quad (1a)$$

$$\mathbf{H}_1(x, y, z) = (E_o/Z_o)(-i\cos\theta\, \hat{\mathbf{x}} + \hat{\mathbf{y}} + i\sin\theta\, \hat{\mathbf{z}})\, \exp[ik_o(x\sin\theta + z\cos\theta)]. \quad (1b)$$

In the above equations, $\hat{\mathbf{x}}$, $\hat{\mathbf{y}}$, and $\hat{\mathbf{z}}$ are Cartesian units vectors, the time-dependence factor $\exp(-i\omega t)$ has been omitted ($\omega = 2\pi f$ is the angular frequency), the propagation direction is along the unit vector $\boldsymbol{\sigma}_1 = (\sin\theta, 0, \cos\theta)$, and the magnitude of the $k$-vector is denoted by $k_o = 2\pi/\lambda_o$, where $\lambda_o = c/f$ is the free-space wavelength. $c = 1/\sqrt{\mu_o \varepsilon_o}$ is the speed of light in vacuum, and $Z_o = \sqrt{\mu_o/\varepsilon_o}$ is the impedance of the free-space [8].

We combine four plane-waves of the type given by Eqs. (1), each propagating in a slightly different direction, as specified by the propagation vectors $\boldsymbol{\sigma}_{1,3} = (\pm\sin\theta, 0, \cos\theta)$ and $\boldsymbol{\sigma}_{2,4} = (0, \pm\sin\theta, \cos\theta)$. The combined $E$-field and $H$-field of the four plane-waves are

$$\boldsymbol{E}(x, y, z) = 2E_o\{[\cos\theta\,\cos(k_o\sin\theta\,x) + \cos(k_o\sin\theta\,y)]\hat{\boldsymbol{x}} + i\,[\cos(k_o\sin\theta\,x) + \cos\theta\,\cos(k_o\sin\theta\,y)]\hat{\boldsymbol{y}}$$
$$- \sin\theta\,[i\sin(k_o\sin\theta\,x) - \sin(k_o\sin\theta\,y)]\hat{\boldsymbol{z}}\}\exp(ik_o\cos\theta\,z), \qquad (2a)$$

$$\boldsymbol{H}(x, y, z) = \frac{2E_o}{Z_o}\{-i\,[\cos\theta\,\cos(k_o\sin\theta\,x) + \cos(k_o\sin\theta\,y)]\hat{\boldsymbol{x}} + [\cos(k_o\sin\theta\,x) + \cos\theta\,\cos(k_o\sin\theta\,y)]\hat{\boldsymbol{y}}$$
$$- \sin\theta\,[\sin(k_o\sin\theta\,x) + i\sin(k_o\sin\theta\,y)]\hat{\boldsymbol{z}}\}\exp(ik_o\cos\theta\,z). \qquad (2b)$$

Plots of the $E$-field intensity and phase distribution in the cross-sectional $xy$-plane for specific values of $\lambda_o$ and $\theta$ are shown in Fig. 1. We are particularly interested in the central lobe (or fringe) of this profile, which occupies a square region of area $A = 2\pi^2/(k_o^2\sin^2\theta)$. In the limit when $\theta \to 0$, this lobe expands to infinity and covers the entire $xy$-plane. The above four-wave superposition allows both $E$- and $H$-fields to maintain a non-zero $z$-component (see, for example, the plot of $E_z$ on the right-hand side of Fig. 1). Ultimately, the non-zero values of $E_z$ and $H_z$ endow the central lobe of the superposition with an angular momentum.

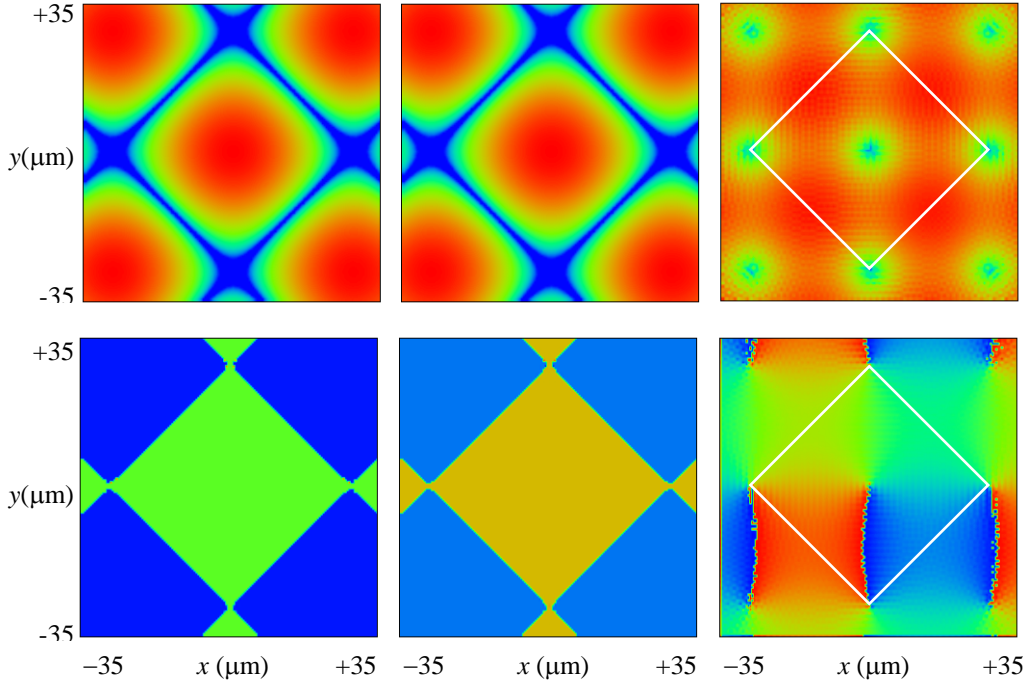

Fig. 1. Plots of $E$-field magnitude (top row) and phase (bottom row) in the $xy$ cross-sectional plane for a superposition of four circularly-polarized plane waves. (Left to right: $E_x$, $E_y$, $E_z$). The assumed wavelength is $\lambda_o = 1.0$ µm, and the four plane-waves are tilted by $\theta = 1°$ relative to the $z$-axis; the central lobe's corners are thus located at $x, y = \pm\pi/(k_o\sin\theta) = 28.65$ µm. A white square is overlaid on the $E_z$ magnitude and phase plots to delineate the central lobe's boundaries. The color coded amplitude plots (top row) are on a logarithmic scale, with the maximum amplitude in each frame coded as red, minimum as blue, and the values in between assigned the rainbow color spectrum. (In each frame, the amplitude plots are truncated at the point where the $E$-field intensity drops below $10^{-3} \times$ Peak_Intensity.). In the $E_x$ and $E_y$ phase plots (bottom row), the central lobe is 180° phase-shifted relative to its four adjacent neighbors. At any given point in the $xy$-plane, $E_y$ is 90° ahead of $E_x$. As for the $E_z$ phase plot (bottom, right), the rainbow colors are assigned such that Blue = 0°, Green = 180°, Red = 360°. In the region of the central lobe (as in all other lobes) $E_z$ shows a vortex-like behavior.

The Poynting vector $S = \tfrac{1}{2}\,\mathrm{Real}\,(\boldsymbol{E}\times\boldsymbol{H}^*)$ can be readily calculated for the four-wave superposition described by Eqs. (2). We find

$$S = 4(E_o^2/Z_o)\{-\sin\theta\,\sin(k_o\sin\theta\,y)\,[\cos(k_o\sin\theta\,x) + \cos\theta\,\cos(k_o\sin\theta\,y)]\,\hat{x}$$
$$+ \sin\theta\,\sin(k_o\sin\theta\,x)\,[\cos\theta\,\cos(k_o\sin\theta\,x) + \cos(k_o\sin\theta\,y)]\,\hat{y}$$
$$+ \{\cos\theta\,[\cos^2(k_o\sin\theta\,x) + \cos^2(k_o\sin\theta\,y)] + (1+\cos^2\theta)\cos(k_o\sin\theta\,x)\cos(k_o\sin\theta\,y)\}\,\hat{z}\} \quad (3)$$

Figure 2 shows plots of $S_x$, $S_y$, $S_z$ for the beam profile of Fig. 1. Note that the central lobe contains, in the $xy$-plane, a circulating component of the Poynting vector around the $z$-axis.

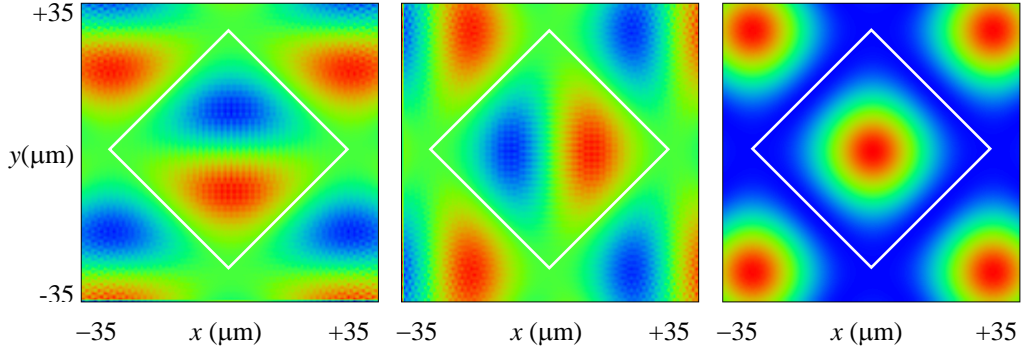

Fig. 2. Left to right: plots of the Poynting vector components $S_x$, $S_y$, $S_z$ in the $xy$ cross-sectional plane of the beam depicted in Fig. 1. (The overlaid white square in each frame delineates the central lobe's boundaries.) In each frame, the maximum value of the depicted function is coded as red, the minimum value as blue, and the values in between are assigned the rainbow color spectrum. The $S_x$ and $S_y$ plots thus show that, in each lobe, the projection of the Poynting vector in the $xy$-plane circulates around the $z$-axis.

The rate of flow of energy through the central lobe is computed by integrating $S_z$ over the (square) region of the lobe. This yields the optical power density $P_o$ (i.e., energy per unit area per unit time) passing through the lobe as follows:

$$<P_o> = [2\pi^2/(k_o^2\sin^2\theta)]^{-1} \iint_{\text{Central lobe}} S_z\,\mathrm{d}x\,\mathrm{d}y = 4(E_o^2/Z_o)\cos\theta. \quad (4)$$

In the limit when $\theta \to 0$ the lobe area extends to infinity and the power density becomes $P_o = 4E_o^2/Z_o$, corresponding to the superposed power density of four circularly-polarized plane waves. Dividing this power density by the vacuum speed of light $c$ yields the electromagnetic field's energy density (i.e., energy per unit volume) in free-space as $4\varepsilon_o E_o^2$, where $\varepsilon_o$ is the permittivity of the free space.

The electromagnetic field's momentum density in free space is $\boldsymbol{p} = \boldsymbol{S}/c^2$; thus the angular momentum density (relative to the origin of the coordinates) is $\boldsymbol{L} = (x, y, z)\times(S_x, S_y, S_z)/c^2$. For the four-wave superposition whose Poynting vector is given by Eq. (3), when $\boldsymbol{L}$ is integrated over the central lobe's area, only the $L_z$ component of the angular momentum survives. After normalization by the lobe area, the average $L_z$ is found to be:

$$<L_z> = 4(E_o^2/Z_o c^2)\,[2\pi^2/(k_o^2\sin^2\theta)]^{-1} \iint_{\text{Central lobe}} \{\tfrac{1}{2}\sin\theta\,\cos\theta\,[x\sin(2k_o\sin\theta\,x) + y\sin(2k_o\sin\theta\,y)]$$
$$+ \sin\theta\,[x\sin(k_o\sin\theta\,x)\cos(k_o\sin\theta\,y) + y\cos(k_o\sin\theta\,x)\sin(k_o\sin\theta\,y)]\}\,\mathrm{d}x\,\mathrm{d}y$$
$$= 4\varepsilon_o E_o^2/(2\pi f). \quad (5)$$

Here $f = c/\lambda_o$ is the oscillation frequency of the electromagnetic field. Note that $<L_z>$ is independent of the inclination angle $\theta$ of the constituent plane-waves. The angular momentum density is thus found to be equal to the energy density $4\varepsilon_o E_o^2$ (in the limit when $\theta \to 0$) divided by $2\pi f$. Considering that the energy of a photon of frequency $f$ is $hf$, where $h$ is Planck's constant, the angular momentum of the photon thus turns out be $\hbar = h/2\pi$, consistent with the principles of quantum mechanics [7].

As an aside, we may use a heuristic argument to assign an effective gyration radius $R_{eff}$ in free-space to a circularly-polarized photon, whose profile we take to be the central lobe of the fringe pattern of Fig. 1. As indicated in Fig. 2, the energy of this photon moves essentially along a spiral in the $z$-direction. The photon's relativistic "mass" is $hf/c^2$, and, because all its constituent $k$-vectors make an angle $\theta$ with $z$, its circulation (i.e., azimuthal) velocity in the $xy$-plane is expected to be $c\sin\theta$. [In fact, according to Eq. (4), the average flux of energy along $z$ is proportional to $\cos\theta$, which leaves the remaining fraction of the energy – that which is proportional to $\sin\theta$ – to circulate.] The azimuthal linear momentum of the photon is thus $p_\phi = hf\sin\theta/c$. The product of $p_\phi$ and $R_{eff}$ is the photon's angular momentum, which is known to be $\hbar$, thus yielding $R_{eff} = 1/(k_o\sin\theta)$. In Fig. 2, where the four corners of the central lobe are at a distance of $\pi/(k_o\sin\theta)$ from the center, the azimuthal component $S_\phi$ of the Poynting vector is seen to be inside the four corners, thus rendering the aforementioned value of $R_{eff}$ entirely plausible. The picture of a circularly polarized photon in free-space that emerges from this discussion is a (more or less compact) bundle of energy spiraling forward with an azimuthal velocity $c\sin\theta$, at an effective distance of $1/(k_o\sin\theta)$ from the $z$-axis. We will use this picture in conjunction with an "Einstein box" argument in Section 6, where we seek to justify the division of angular momentum within dielectric media into electromagnetic and mechanical parts. (Needless to say, our naïve classical picture of a photon should not be taken too seriously in a quantum world.)

## 3. Four-wave superposition inside a dielectric medium

We now consider the case of four superposed, circularly-polarized plane-waves inside a dielectric medium of refractive index $n$, as depicted schematically in Fig. 3. Initially, we assume a complex dielectric constant $\varepsilon = (n + i\kappa)^2$ for the medium, where $\kappa$ is the absorption coefficient. Eventually, however, $\kappa$ will be allowed to approach zero (and thus $\varepsilon$ to become real-valued) in the limit. Crossing the $xy$-plane at $z = 0$, the four plane-waves enter the semi-infinite dielectric from the free-space. Since the angle of incidence $\theta$ is small (and will eventually go to zero), the reflection coefficient at the free-space-to-dielectric interface is assumed to be $r = (1 - \sqrt{\varepsilon})/(1 + \sqrt{\varepsilon})$, independent of the incident beam's polarization state. (The dependence of $r$ on $\kappa$, however, should *not* be ignored at the outset, even though $\kappa$ will eventually go to zero.) In analogy with Eqs. (2), but with the dielectric constant $\varepsilon$ and the reflection coefficient $r$ taken into consideration, the $E$- and $H$-fields within the dielectric are

$$\boldsymbol{E}(x, y, z) = 2E_o\{(1+r)[\cos\theta\cos(k_o\sin\theta\, x) + \cos(k_o\sin\theta\, y)]\hat{\boldsymbol{x}}$$
$$+ \mathrm{i}(1+r)[\cos(k_o\sin\theta\, x) + \cos\theta\cos(k_o\sin\theta\, y)]\hat{\boldsymbol{y}}$$
$$- (1-r)\varepsilon^{-1}\sin\theta[\mathrm{i}\sin(k_o\sin\theta\, x) - \sin(k_o\sin\theta\, y)]\hat{\boldsymbol{z}}\}\exp(\mathrm{i}k_o\sqrt{\varepsilon - \sin^2\theta}\, z), \quad (6a)$$

$$\boldsymbol{H}(x, y, z) = 2(E_o/Z_o)\{-\mathrm{i}(1-r)[\cos\theta\cos(k_o\sin\theta\, x) + \cos(k_o\sin\theta\, y)]\hat{\boldsymbol{x}}$$
$$+ (1-r)[\cos(k_o\sin\theta\, x) + \cos\theta\cos(k_o\sin\theta\, y)]\hat{\boldsymbol{y}}$$
$$- (1+r)\sin\theta[\sin(k_o\sin\theta\, x) + \mathrm{i}\sin(k_o\sin\theta\, y)]\hat{\boldsymbol{z}}\}\exp(\mathrm{i}k_o\sqrt{\varepsilon - \sin^2\theta}\, z). \quad (6b)$$

As before, the four-wave superposition in Eqs. (6) creates a stationary fringe pattern in the $xy$-cross-sectional plane of the beam. Of primary concern, of course, will be the central lobe of this fringe pattern, which is similar to that shown in Fig. 1. We mention in passing that the beams described by either Eqs. (2) or Eqs. (6) are diffraction-free beams. (The fact that all the constituent *k*-vectors make the same angle θ with the *z*-axis ensures diffraction-free behavior.) The only effect of propagation on the fringe pattern, evident in Eqs.(2) and (6), is a constant phase factor – the exponential term at the end of the expressions for *E*- and *H*-fields – that multiplies the beam's cross-sectional profile.

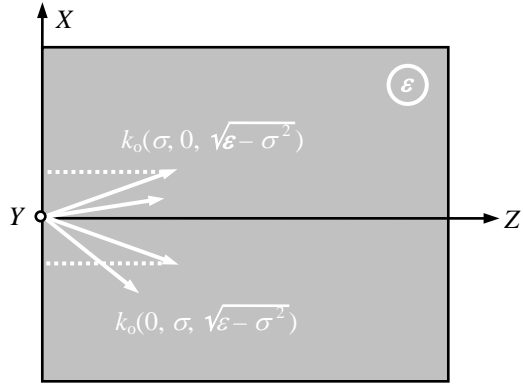

Fig. 3. Four circularly polarized plane-waves enter a medium of dielectric constant $\varepsilon$. The *k*-vectors of the plane-waves are $k_{1,3} = k_o(\pm\sigma, 0, \sqrt{\varepsilon - \sigma^2})$ and $k_{2,4} = k_o(0, \pm\sigma, \sqrt{\varepsilon - \sigma^2})$. Here $\sigma = \sin\theta$, where θ is the angle of incidence in the free space.

We use a two-step procedure to compute the forces and torques exerted by the central lobe of the fringe pattern on the dielectric medium. In the first step, elaborated in Section 4, the bound electrical charge density $\sigma_b(x, y, z=0)$, induced at the entrance facet of the semi-infinite dielectric, is computed, and the Lorentz force of the local *E*-field on these charges is determined. In addition to an outward-directed force, the surface charge distribution will be found to produce a net torque $\boldsymbol{T}^{(surface)}$ along *z*. (Since the imaginary part $\kappa$ of $\sqrt{\varepsilon}$ does not play a critical role as far as surface forces are concerned, it will be set to zero at the outset.)

In the second step, described in Section 5, we compute the Lorentz force of the *H*-field on the bound electrical currents $\boldsymbol{J}_b(x, y, z)$, induced by the light beam's *E*-field in the bulk of the dielectric. These forces also produce a net torque $\boldsymbol{T}^{(bulk)}$ along the *z*-axis. The imaginary part $\kappa$ of $\sqrt{\varepsilon}$ turns out to be crucial in these calculations because, in the absence of $\kappa$, the magnetic Lorentz force vanishes everywhere, whereas with $\kappa \neq 0$ the Lorentz force acquires a non-zero value. As $\kappa$ goes from a finite value to zero, the density of the magnetic Lorentz force in the bulk decreases but, at the same time, the light penetrates deeper in the medium. The net result is that the integrated force and torque now approach a definite non-zero value in the limit when $\kappa \to 0$. In other words, the extended range of integration (due to a deeper penetration of light) compensates for the weakness of the localized force and/or torque when $\kappa$ is small.

## 4. Surface contributions to force and torque

The discontinuity of $E_z$ at the interface between the free-space and the dielectric medium produces a surface density of bound charges, given by $\sigma_b = \varepsilon_o \Delta E_z$. The magnitude of $E_z$ just inside the dielectric (i.e., beneath the surface) is given in Eq. (6a), and the magnitude of $E_z$ just above the surface is readily obtained from the continuity of the perpendicular component of the displacement vector, $D_z = \varepsilon_o \varepsilon E_z$. We find

$$\sigma_b(x, y, z=0) = 2\varepsilon_o E_o(1-r)(1-\varepsilon^{-1})\sin\theta \, [i\sin(k_o\sin\theta \, x) - \sin(k_o\sin\theta \, y)]. \qquad (7)$$

This surface charge interacts with the local *E*-field to produce a surface force density $\boldsymbol{F} = \frac{1}{2}Real(\sigma_b \boldsymbol{E}^*)$. Since the tangential *E*-field is continuous across the boundary, $E_x$ and $E_y$ evaluated from Eq. (6a) at $z=0$ can be used to compute the tangential force components $F_x$, $F_y$ at the surface. As for the perpendicular component $F_z$, however, the discontinuity of $E_z$ at the boundary requires that the average of $E_z$ be used for force computation [9]. Setting $\varepsilon = n^2$ and $r = (1-n)/(1+n)$, we find

$$F_z(x, y, z=0) = -\frac{4(n-1)(n^2+1)\varepsilon_o E_o^2}{n^2(n+1)} \sin^2\theta \left[\sin^2(k_o\sin\theta\, x) + \sin^2(k_o\sin\theta\, y)\right]. \tag{8}$$

When integrated over the area of the central lobe, the total force along the z-axis becomes

$$\iint_{\text{Central lobe}} F_z(x, y, z=0)\,dx\,dy = -\frac{2(n-1)(n^2+1)}{n^2(n+1)} \varepsilon_o E_o^2 \lambda_o^2 = -\frac{(n-1)(n^2+1) <P_o> \lambda_o^2}{2n^2(n+1)\,c}, \tag{9}$$

where $<P_o>$ is the average power density crossing the lobe; see Eq. (4). This force is independent of $\theta$ and, therefore, its density (i.e., force normalized by the lobe area $A$) diminishes as $\theta \to 0$. The strength of the z-component of the surface force, given by Eq. (9), does not depend on the lobe area $A$, so long as $E_o$ is treated as a constant. If, however, a beam of total power $Q_o$ and foot-print area $A$ is normally incident on a dielectric, then the total force on the surface, in accordance with Eq. (9), will be proportional to $<P_o>\lambda_o^2 = Q_o/(A/\lambda_o^2)$. The normalized footprint $A/\lambda_o^2$ of the beam in this case plays an important role in determining the overall strength of the surface force. We believe this surface force is responsible for the bulge of the water surface observed in the experiments of Ashkin and Dziedzic [13]. With a laser beam having $Q_o = 1.0$ kW, focused to an area $A = 100\lambda_o^2$ on the surface of water ($n = 1.335$ at $\lambda_o = 0.53\,\mu$m), Eq. (9) predicts a net force of $\sim 3.7$ nN, which would have been sufficient to lift the water in Ashkin's experiment by a few microns over a millimeter-sized area.

In any event, the symmetry of the $F_z$ distribution in Eq. (8) precludes a torque around the center of coordinates. In the following discussion, therefore, we concentrate on the tangential force $F_x\hat{x} + F_y\hat{y}$. Anticipating the limit operation where $\kappa$ and $\theta$ will approach zero, we set $\varepsilon = n^2$, $r = (1-n)/(1+n)$, and $\cos\theta = 1$. The tangential force density is then given by

$$F_x\hat{x} + F_y\hat{y} = \frac{8(n-1)\varepsilon_o E_o^2}{n(n+1)} \sin\theta \left[\cos(k_o\sin\theta\, x) + \cos(k_o\sin\theta\, y)\right]\left[-\sin(k_o\sin\theta\, y)\hat{x} + \sin(k_o\sin\theta\, x)\hat{y}\right] \tag{10}$$

The average torque experienced by the dielectric surface due to the (bound) surface charges $\sigma_b$ may thus be found by integrating the torque density $T_z = xF_y - yF_x$ over the central lobe, then normalizing by the lobe's area, that is,

$$<T_z^{(\text{surface})}> = \frac{8(n-1)\varepsilon_o E_o^2}{n(n+1)} [2\pi^2/(k_o^2\sin^2\theta)]^{-1} \times \sin\theta \iint_{\text{Central lobe}} [\tfrac{1}{2}x\sin(2k_o\sin\theta\, x) + \tfrac{1}{2}y\sin(2k_o\sin\theta\, y)$$
$$+ x\sin(k_o\sin\theta\, x)\cos(k_o\sin\theta\, y) + y\cos(k_o\sin\theta\, x)\sin(k_o\sin\theta\, y)]\,dx\,dy$$
$$= \frac{8(n-1)\varepsilon_o E_o^2}{n(n+1)k_o}. \tag{11}$$

A fraction of the incident angular momentum is thus transferred as torque to the dielectric surface. We emphasize the important difference between the surface torque of Eq. (11) and the surface force of Eq. (9). The force in Eq. (9) is the total force, *not* the force per unit area; therefore, for a given power density $<P_o>$, if the cross-sectional area $A$ of the beam is increased, the force per unit area will decrease, even though the total force remains constant.

In contrast, the torque given by Eq. (11) is torque per unit surface area, which remains the same for all values of $A$, so long as the power density $<P_o>$ is kept constant.

## 5. Bulk contributions to force and torque

The next step involves the calculation of the net force and torque exerted on the bulk of the dielectric by the light that crosses the surface and enters the medium; the relevant $E$- and $H$-fields are given by Eqs. (6). The magnetic field contribution to the Lorentz force, $\boldsymbol{F} = \boldsymbol{J}_b \times \boldsymbol{B}$, is the only force at work here, as the electric field contribution, $\boldsymbol{F} = \rho_b \boldsymbol{E}$, is exactly zero by virtue of the fact that the bound charge density $\rho_b = -\nabla \cdot \boldsymbol{P} = -\varepsilon_o(\varepsilon - 1)\nabla \cdot \boldsymbol{E} = (\varepsilon^{-1} - 1)\nabla \cdot \boldsymbol{D} = 0$. Since the bound current density $\boldsymbol{J}_b = \partial \boldsymbol{P}/\partial t$, where $\boldsymbol{P} = \varepsilon_o(\varepsilon - 1)\boldsymbol{E}$, we have, for small $\theta$ and $\kappa$,

$$\boldsymbol{F}(x,y,z) = \tfrac{1}{2} \mathrm{Real}(\boldsymbol{J}_b \times \boldsymbol{B}^*) = \tfrac{1}{2} \mathrm{Real}[-i\omega\varepsilon_o(\varepsilon - 1)\boldsymbol{E} \times \mu_o \boldsymbol{H}^*]$$

$$\approx \frac{16 k_o \varepsilon_o E_o^2}{(n+1)^2} \Big\{ (n+n^{-1}) \sin\theta \left[\cos(k_o \sin\theta\, x) + \cos(k_o \sin\theta\, y)\right] \left[-\sin(k_o \sin\theta\, y)\hat{\boldsymbol{x}} + \sin(k_o \sin\theta\, x)\hat{\boldsymbol{y}}\right]$$
$$+ (n^2+1)[\cos(k_o \sin\theta\, x) + \cos(k_o \sin\theta\, y)]^2 \hat{\boldsymbol{z}} \Big\} \kappa \exp(-2 k_o \kappa z). \tag{12}$$

In the above equation, terms of second- and higher-order in $\kappa$ and $\sin\theta$ have been ignored; in particular, $\cos\theta$ has been set equal to unity.

In the $z$-direction the force drops exponentially away from the entrance surface. If we set $\kappa$ equal to zero (to represent a pure dielectric), the entire force vanishes. However, if the last term in Eq. (12), $\kappa \exp(-2 k_o \kappa z)$, is integrated from $z=0$ to infinity, it reduces to $1/(2 k_o)$, which is independent of $\kappa$. In the limit $\kappa \to 0$, therefore, the total force on the bulk of a semi-infinite dielectric turns out to have a finite (non-zero) value.

The $z$-component of the force $\boldsymbol{F}$ in Eq. (12) does *not* produce a torque around the $z$-axis. However, its average over the central lobe, integrated from $z=0$ to $\infty$, is evaluated as follows:

$$<F_z^{(bulk)}> = [2\pi^2/(k_o^2 \sin^2\theta)]^{-1} \underset{\text{Central lobe}}{\iint} \mathrm{d}x\,\mathrm{d}y \int_0^\infty F_z(x,y,z)\,\mathrm{d}z = \frac{8(n^2+1)}{(n+1)^2} \varepsilon_o E_o^2 \tag{13}$$

Considering that the influx of the linear momentum in the incident beam is $4\varepsilon_o E_o^2$ per unit area per unit time, and that the factor $1 + |r|^2 = 2(n^2+1)/(n+1)^2$ is needed to account for the total momentum transfer from the incident and reflected beams to the dielectric medium, we see that, in the limit when $\theta \to 0$, Eq. (13) yields the correct expression for the force density (i.e., force per unit surface area) exerted on the entire body of the semi-infinite dielectric.

The angular momentum density experienced by the bulk of the dielectric may now be derived from the $x$- and $y$-components of $\boldsymbol{F}$ in Eq. (12) by integration and normalization, i.e.,

$$<T_z^{(bulk)}> = [2\pi^2/(k_o^2 \sin^2\theta)]^{-1} \underset{\text{Central lobe}}{\iint} \Big\{ x \int_0^\infty F_y(x,y,z)\,\mathrm{d}z - y \int_0^\infty F_x(x,y,z)\,\mathrm{d}z \Big\} \mathrm{d}x\,\mathrm{d}y$$

$$= \frac{8(n+n^{-1})\varepsilon_o E_o^2}{k_o(n+1)^2} \tag{14}$$

Adding Eqs. (11) and (14) yields the total torque per unit surface area as follows:

$$<T_z^{(surface)} + T_z^{(bulk)}> = \frac{16 n \varepsilon_o E_o^2}{k_o(n+1)^2} = [4\varepsilon_o E_o^2 c/(2\pi f)](1 - |r|^2). \tag{15}$$

This is exactly equal to the rate of arrival of angular momentum contained in the incident beam, minus its rate of departure, as the reflected light takes away some of the incident angular momentum. Considering that the sense of circular polarization is reversed upon reflection (i.e., right-circular becomes left-circular, and vice-versa), the failure of the reflection process to transfer any torque to the dielectric should not be surprising. (This is unlike the situation with linear momentum, where a force is exerted on the medium in the process of reflection.) With the reflected light properly discounted, we conclude that the rate of arrival of angular momentum at the surface equals the torque exerted on the dielectric in its entirety (i.e., surface and bulk), provided that the medium is sufficiently thick to eventually absorb all the optical energy that is transmitted from the free-space into the dielectric.

Returning now to Eq. (14), we denote by $\tau$ the $E$-field's amplitude transmission coefficient at the surface, that is, $\tau = 1 + r = 2/(1+n)$. The rate of flow of angular momentum into the bulk of the semi-infinite dielectric (excluding the front surface) must be equal to the torque $<T_z^{(bulk)}>$ given by Eq. (14). [Note that, by the time we arrived at Eq. (14), the fraction of the incident angular momentum reflected at the surface as well as the fraction given as torque to the surface charges were already removed from the incident angular momentum.] Thus, dividing $<T_z^{(bulk)}>$ by the speed of light $c/n$ in the (dispersionless) dielectric should yield the (volume) density $L_z$ of angular momentum inside the medium, that is,

$$L_z = \tfrac{1}{2}(1+n^{-2})\left[4\varepsilon_o \varepsilon (\tau E_o)^2/(2\pi f)\right]. \tag{16}$$

Equation (16) is our final expression for the density of angular momentum inside a transparent dielectric of refractive index $n = \sqrt{\varepsilon}$. Note that the electromagnetic energy density of the four-wave beam inside the medium is $4\varepsilon_o \varepsilon(\tau E_o)^2$. Therefore, the angular momentum of a photon of energy $hf$ inside the dielectric should be $\tfrac{1}{2}(1+n^{-2})\hbar$. This is somewhat lower than the photon's initial angular momentum of $\hbar$ in free-space because, as discussed in Section 4, crossing into the dielectric converts a fraction of the incident angular momentum into a torque exerted on the induced charges at the dielectric surface.

We have argued in [9] that a photon's linear momentum inside a dielectric has the average of its Minkowski and Abraham values, namely, $p = \tfrac{1}{2}(n + n^{-1})hf/c$. In our four-wave picture of a photon, the $k$-vectors of the constituent plane-waves, upon crossing from the free-space into the dielectric, bend toward the $z$-axis in accordance with Snell's law ($\sin\theta' = n^{-1}\sin\theta$), thus reducing (by a factor of $n$) the azimuthal component of the momentum (i.e., the projection of $\boldsymbol{p}$ in the $xy$-plane). All in all, the photon's angular momentum in the dielectric is expected to have its free-space value of $\hbar$ multiplied by $\tfrac{1}{2}(n+n^{-1})/n$, which is exactly what the preceding analysis has indicated. The circularly-polarized photon inside the dielectric thus has the average of its Minkowski ($\hbar$) and Abraham ($\hbar/n^2$) angular momenta.

The photon's linear momentum $p$ inside a dielectric was shown in [9] to be divided into an electromagnetic part, $hf/(nc)$, and a mechanical part, $\tfrac{1}{2}(n - n^{-1})hf/c$. By the same token, we now recognize the photon's angular momentum inside the dielectric as consisting of an electromagnetic part, $\hbar/n^2$, and a mechanical part, $\tfrac{1}{2}(1 - n^{-2})\hbar$. The photon's mechanical momenta (both linear and angular) are part and parcel of the same wave-packet, dressing the electromagnetic portion of the packet with mechanical motions of the constituent atoms of the dielectric, and (ignoring acoustic effects) travel alongside the packet through the entire medium. When the photon leaves the slab at the exit facet, these mechanical momenta are converted back into their electromagnetic counterparts, leaving a motionless medium behind.

## 6. Summary and discussion

We have shown the angular momentum density of a circularly-polarized plane-wave inside a dispersionless dielectric of refractive index $n$ is reduced by a factor of $\tfrac{1}{2}(1+n^{-2})$ from the value that one would expect based on a simple conservation of angular momentum argument. The difference, namely, a fraction $\tfrac{1}{2}(1 - n^{-2})$ of the available angular momentum, is given as torque to the (bound) electrical charges induced on the dielectric surface upon the beam's entry. If the medium is in the form of a slab, then an equal but opposite torque will be exerted on the exit facet when the beam leaves the dielectric medium. The end result is that the slab will pick up no net torque, and the light beam will recover its original angular momentum (aside from reflection losses at the two facets) when it returns to the free-space.

The beam's angular momentum inside the dielectric consists of both electromagnetic and mechanical contributions. After removing the fraction of light that is reflected at the surface (this fraction does not give any angular momentum to the dielectric), the fraction of the remaining (i.e., transmitted) angular momentum that is electromagnetic in nature is $1/n^2$, while the fraction that is mechanical is $½(1-n^{-2})$. In addition to this mechanical momentum, another (equal) fraction, $½(1-n^{-2})$, of the available angular momentum is transferred to the dielectric surface at the time of entry. The total fraction of mechanical angular momentum given to the dielectric (surface plus bulk) is thus $(1-n^{-2})$. This result is in agreement with the results of Padgett *et al* [14], except for our division of the mechanical angular momentum into equal contributions from the bulk and the surface.

Our results differ somewhat from those obtained by Loudon [7] in that his reduction factor for the angular momentum inside the dielectric is $1/n^2$, with the remaining fraction, $(1-n^{-2})$, converted to torque at the entrance facet. We suspect that our differences can be traced to the different formulas used to compute the *E*-field's contribution to the Lorentz force: whereas we use $\boldsymbol{F} = -(\boldsymbol{\nabla}\cdot\boldsymbol{P})\boldsymbol{E} + (\partial\boldsymbol{P}/\partial t)\times\boldsymbol{B}$, Loudon's starting point is $\boldsymbol{F} = (\boldsymbol{P}\cdot\boldsymbol{\nabla})\boldsymbol{E} + (\partial\boldsymbol{P}/\partial t)\times\boldsymbol{B}$; he also adds the term $\boldsymbol{P}\times\boldsymbol{E}$ to the torque density formula $\boldsymbol{T} = \boldsymbol{r}\times\boldsymbol{F}$. We have discussed in [9] the merits and demerits of the two versions of the Lorentz force in the context of dense and rare dielectrics. Experimental data are needed to settle the question as to which formulation leads to more reliable predictions.

Padgett *et al* [14] have rightly emphasized the importance of Einstein box (gedanken) experiments in conjunction with the problems of linear and angular momenta of light in dielectrics. The picture of a circularly-polarized photon described at the end of Section 2 may now be used to verify the consistency of our results with an Einstein box type of experiment. Consider a dielectric slab of thickness *d* and refractive index *n* in free-space. The travel time for a photon of angular momentum $\hbar$ through this slab is $t = nd/c$. If the photon traveled in free-space for the same length of time, its integrated angular momentum would be $nd\hbar/c$. In the dielectric slab, however, the electromagnetic part of the photon's angular momentum is reduced by a factor of $n^2$. [This is because the effective radius $R_{eff}$ of the photon remains the same as in free-space, while the inclination angles θ of the constituent plane-waves drop to θ′ (Snell's law), and also because the speed of light in the medium is reduced, leading to a reduction of the photon's azimuthal speed from $c\sin\theta$ to $(c/n)\sin\theta'$.] The integrated angular momentum of the photon for the duration spent inside the slab is thus $d\hbar/(nc)$. While the photon is inside, the slab acquires the mechanical part (surface plus bulk) of the angular momentum, $(1-n^{-2})\hbar$, whose time integral is $(n-n^{-1})d\hbar/c$. (When divided by the slab's moment of inertia, this integrated angular momentum yields the rotation angle of the slab around the *z*-axis, which should be a measurable quantity.) The sum of the integrated momenta of the slab and the transmitted photon thus equals the integrated momentum of the photon traveling for the same duration of time in free-space, as it should. Extension to the case of dispersive media where the refractive index $n_g$ (for group velocity) differs from $n_p$ (used in Snell's law) is straightforward and leads to the same result as given in [14].

**Acknowledgments**

The author is grateful to Ewan Wright, Armis Zakharian, Pavel Polynkin, Rodney Loudon, and Miles Padgett for helpful discussions. This work is supported by the AFOSR contract F49620-02-1-0380 with the Joint Technology Office, by the *Office of Naval Research* MURI grant No. N00014-03-1-0793, and by the *National Science Foundation* STC Program under agreement DMR-0120967.